\documentclass[aps,prl,nofootinbib,preprintnumbers,amsmath,amssymb,latexsym,array,enumerate,letter,twocolumn,superscriptaddress]{revtex4}
\usepackage{amsmath}
\usepackage{color}
\usepackage{graphicx}
\usepackage{slashed}
\usepackage{tcolorbox}
\usepackage{ulem}

\def\a{\alpha}
\def\b{\beta}
\def\c{\gamma}

\def\D{\Delta}
\def\e{\varepsilon}

\def\w{\omega}

\def\La{\mathcal{L}}

\def\bigO{\mathcal{O}}
\def\p{\partial}

\newcommand{\mc}{\mathcal}

\newcommand{\uu}{\;\!}

\begin{document}

\title{Dark Photon Polarimetry}

\author{Fiesta Ting Yan Leung}
\email{ftyleung@connect.ust.hk}
\affiliation{Department of Physics and Jockey Club Institute for Advanced Study, The Hong Kong University of Science and Technology, Hong Kong S.A.R., China}

\author{Tao Liu}
\email{taoliu@ust.hk}
\affiliation{Department of Physics and Jockey Club Institute for Advanced Study, The Hong Kong University of Science and Technology, Hong Kong S.A.R., China}

\author{Sida Lu}
\email{lusida1992@gmail.com}
\affiliation{Institute for Advanced Study, The Hong Kong University of Science and Technology, \\Clear Water Bay, Kowloon, Hong Kong S.A.R., China}

\author{Jing Ren}
\email{renjing@ihep.ac.cn}
\affiliation{Institute of High Energy Physics, Chinese Academy of Sciences, Beijing 100049, China}	
\affiliation{Center for High Energy Physics, Peking University, Beijing 100871, China}

\author{Cheuk Kan Kelvin Yue}
\email{ckkyue@connect.ust.hk}
\affiliation{Department of Physics and Jockey Club Institute for Advanced Study, The Hong Kong University of Science and Technology, Hong Kong S.A.R., China}

\author{Kaifeng Zheng}
\email{kzhengae@connect.ust.hk}
\affiliation{Department of Physics and Jockey Club Institute for Advanced Study, The Hong Kong University of Science and Technology, Hong Kong S.A.R., China}

\begin{abstract}
We propose detecting dark photons (DP), a major candidate for wave dark matter, through polarimetry. The DP can modify Maxwell's equations due to its kinetic mixing with regular photons, inducing an oscillating component in the electromagnetic field. This may leave an imprint on polarimetric light signals, such as Faraday rotation, characterized by a distinctive wave pattern in spacetime. We then apply this method to investigate ultralight DPs produced through the superradiance of supermassive black holes for demonstration. Using the polarimetric measurements of radiation from the M87$^\ast$ and Sgr A$^*$ at the Event Horizon Telescope, we show that novel limits on the photon-DP mixing parameter can be set for the rarely-explored DP mass ranges: explicitly $10^{-22} - 10^{-20}\,$eV with the best reach of $\sim 10^{-8}$ achieved at $\sim 10^{-20}$eV, and $10^{-19} - 10^{-17}\,$eV with the best reach of $\sim 10^{-11}$ achieved at $\sim 10^{-17}$eV. Given the universality of its underlying physics, we expect DP polarimetry to be broadly applicable for DP detection in laboratory experiments and astronomical observations.

\end{abstract}

\maketitle

\section{Introduction}
Despite compelling evidence~\cite{Zwicky:1933gu,Planck:2018vyg,Markevitch:2003at} for their existence, the nature of dark matter (DM) remains elusive. The null results from direct DM detections~\cite{XENON:2023cxc,PandaX-4T:2021bab,LZ:2022lsv} and complementary searches at the Large Hadron Collider challenge the conventional model of weakly interacting massive particles, where the DM mass is typically at electroweak scale, advocating for alternative models. Among the suggested possibilities, bosonic particles with a mass of $\lesssim 1$eV emerge as an intriguing DM candidate. With their wave nature overshadowing their particle characteristics, these bosons are often referred to as ``wave DM'' (for a review, see, e.g.,~\cite{Hui:2021tkt}) and possess unique astrophysical and cosmological implications~\cite{Redondo:2013lna,Bar:2019bqz,Schive_2014,Hu:2000ke,Hlozek:2017zzf,Nori:2018pka,Braaten:2015eeu}.

One representative example of wave DM is dark photon (DP)~\cite{Holdom:1985ag}, a massive Abelian vector boson ($A'$) that interacts with the Standard Model of particle physics through a kinetic mixing with the regular photon ($A$), described by the Lagrangian $\La\supset -\frac{\e}{2}F^\prime_{\mu\nu}F^{\mu\nu}$. Here, $F^\prime_{\mu\nu}$ and $F^{\mu\nu}$ represent the field strengths of the DP and the photon, and $\varepsilon$ is their kinetic mixing parameter. In last decades, extensive experiments and observations have been taken and many novel methods have been proposed for investigating the DP non-gravitational properties~\cite{Bahre:2013ywa,DarkSide:2022knj,XENON:2021qze,Cervantes:2022epl,An:2022hhb,Caputo:2020bdy,Wadekar:2019mpc,Yan:2023kdg}. These efforts primarily leverage mixing-induced DP-photon conversions and DP interactions with charged particles, yielding stringent constraints on its mixing parameter across a wide mass range (see, e.g.,~\cite{Caputo:2021eaa}, for a summary of existing limits). However, for the mass regime of ultralight wave DM ($\lesssim 10^{-18}$eV)~\cite{Marsh:2015xka}, including the well-known fuzzy DM scenario~\cite{Hu:2000ke,hui2017ultralight} which offers a potential solution to the ``small-scale'' structure problems, this mixing remains rarely explored.

To address this challenge, in this Letter we propose detecting the DP through polarimetry. This optical methodology has been broadly utilised to detect axions or axion-like particles, another prominent candidate for wave DM. When linearly polarised light travels across an axion field or DM halo, its polarisation position angle (PA) can be rotated due to a topological effect induced by the axion Chern-Simons coupling. This phenomenon is commonly referred to as ``cosmic birefringence (CB)''~\cite{Carroll:1989vb,Carroll:1991zs}. Over the past decades, various polarimetric observation tools or methods have been developed to measure this effect, including  Cosmic Microwave Background (CMB)~\cite{Lue:1998mq,Feng:2006dp,Fedderke:2019ajk}, pulsar polarisation arrays~\cite{Liu:2021zlt,Xue:2024zjq}, supermassive black holes (BHs)~\cite{chen2021stringent,yuan2021testing,chen2022birefringence,Gan:2023swl}, and Crab Nebula~\cite{Castillo:2022zfl,POLARBEAR:2024vel}, etc.

In contrast to the case of the axion or axion-like DM, the DP polarimetry is based on the effective current in the Maxwell's equations which arises from the photon-DP kinetic mixing. This effect induces an oscillating component in the electromagnetic (EM) field. In magneto-optic media ({\it e.g.}, magnetized plasma), it may leave an imprint in the polarimetric light signals such as Faraday rotation (FR) and Faraday conversion (FC), characterised by a specific wave pattern in spacetime. Given the universality of its underlying physics, the DP polarimetry can be applied for the DP detection in both laboratory experiments and astronomical or cosmological observations. For demonstration, below we will consider supermassive Kerr (or spinning) BHs surrounded by dense thermal plasma as a laboratory for performing this investigation.

The ultralight DP is known to be subject to superradiance near a Kerr BH, by draining its energy and angular momentum. When their Compton wavelength is comparable to the BH horizon, requiring the BH to be supermassive, a macroscopic cloud of DPs can be exponentially populated, forming a coherent state. The cloud profiles are characterised by a set of quantum numbers $(n,j,l,m)$, similar to electron eigenstates of a hydrogen atom. Thus, such a system is often dubbed ``gravitational atom''~\cite{Detweiler:1980uk}. The  ultrlight DP can be sought through polarimetric observations of the supermassive BHs such as those at the centers of the galaxy Messier 87 (M87$^\ast$) and the Milky Way (Sagittarius (Sgr) A$^*$) using the Event Horizon Telescope (EHT)~\cite{EHT-M87-1,EventHorizonTelescope:2022wok}. Below, while emphasizing the role of the FR observable in constraining the photon-DP kinetic mixing, we will show that other Stokes parameters could also serve as a sensible probe for the DP.

\section{Dark Photon Polarimetry and Black Holes}
The Maxwell's equations with a photon-DP kinetic mixing are given by
\begin{align}
    \p_{\mu}F^{\mu\nu}=-(J^\nu+J^{\prime\nu})\, .
    \label{eq:Max}
\end{align}
Here, $J^\nu$ denotes the regular EM current, and $J^{\prime\nu}=-\e\mu^{2}A^{\prime\nu}+\bigO(\e^2)$ is induced by the photon-DP mixing, with $\mu$ being the DP mass parameter. Since these equations are linear, the EM field strength components sourced by $J^{\prime\nu}$ can be isolated from the others when studying the DP polarimetry.

For gravitational atoms with a DP cloud, the DP field in Eq.~(\ref{eq:Max}) should be interpreted as the cloud profile. To the leading order of gravitational fine-structure constant, namely $\a=GM\mu$ ($G$ is the gravitational constant and $M$ is the BH mass), the profile for the fastest-growing mode $(n,j,l,m)=(1,1,0,1)$ is given by~\cite{baryakhtar2017black,Siemonsen:2022ivj}
\begin{equation}
    \begin{aligned}
        A^\prime_0&=\dfrac{\sqrt{M_c}}{\sqrt{\pi}\mu^2 r^{5/2}_c}e^{-r/r_c}\sin\theta\sin(\w t-\varphi+\phi_0) \, ,\\
         \vec{A}^\prime&=-\frac{\sqrt{M_c}}{\sqrt{\pi}\mu\uu r^{3/2}_c}e^{-r/r_c}\{\cos(\w t +\phi_0),\sin(\w t+\phi_0),0\} \, .
    \end{aligned}
    \label{eq:superrad_profile}
\end{equation}
Here, $r$ represents ``$r-r_+$'' in the Boyer-Lindquist coordinate system, with $r_+$ being the Kerr BH horizon, $\theta$ is  polar angle, defined with respect to the BH spinning axis, and $\varphi$ is azimuthal angle. Additionally, $r_{c}=r_{g}/\a^2=1/(GM\mu^2)$ is gravitational Bohr radius, with $r_{g}= GM$ representing half Schwarzschild radius, $M_c\sim0.1\a M$ is DP cloud mass for $\a \ll 1$, $\w\simeq\mu(1-\a^2/2)$ is DP eigenenergy, and $\phi_0$ is a random constant phase. We will take this profile below for this proof-of-concept study.  

Given the $J^{\prime\nu}$, the DP-induced EM field strength is then solved to be 
\begin{equation}
    \begin{aligned}
        \vec{B}_{A^\prime}(t,\vec{x})&=\int dt^\prime d\vec{x}^\prime G(t,\vec{x};t^\prime,\vec{x}^\prime)\left(\nabla\times \vec{J}'(t^\prime,\vec{x}^\prime)\right) \, ,
        \label{eq:B_DP}\\
        \vec{E}_{A^\prime}(t,\vec{x})&=-\int dt^\prime d\vec{x}^\prime G(t,\vec{x};t^\prime,\vec{x}^\prime)\left(\nabla J'^0 + \partial_{t^\prime}\vec{J}'(t^\prime,\vec{x}^\prime)\right)\, .
    \end{aligned}
\end{equation}
Here, $G(t,\vec{x};t^\prime,\vec{x}^\prime)$ is Green's function (GF), with $(t^\prime, \vec{x}^\prime)$ and $(t,\vec{x})$ denoting the positions of the EM source and field, respectively. As $A'_0\sim\a|\vec{A}'|$ and $1/r_c\sim\a\mu$, the contribution of $\nabla J'^0$ to $\vec{E}_{A^\prime,i}(t,\vec{x})$ can be neglected. 

A rigorous treatment of the GF requires summing the eigenfunctions in the Kerr metric after applying separation of variables, as illustrated in~\cite{Teukolsky:1972my}. However, the signal signature primarily forms on the BH accretion disk, whose inner edge is located outside the photon sphere and not very close to the event horizon ($r_+\sim r_{g}$). The metric in this area differs from that of flat spacetime only up to a factor of $\sim \bigO(1)$. Therefore, we will take the GF in the flat spacetime as a leading-order approximation in this exploratory study. Moreover, the current $J^{\prime\nu} \propto A^{\prime\nu}$ is exponentially suppressed in the region of $r>r_{c}$, so we expect the PA polarimetry to be particularly relevant for $r_g < r\ll r_c$. In this context, the DP-induced EM fields are approximately given by:
\begin{equation}
    \begin{aligned}
        B_{A^\prime,r}&= 0 \,  ,\\
        B_{A^\prime,\theta}&\approx0.2 \e\a^{11/2}\Lambda\left(\frac{r}{r_{g}}\right)\cos(\w t-\varphi+\phi_{0}) \,,\\
        B_{A^\prime,\varphi}&\approx0.2\e\a^{11/2}\Lambda\left(\frac{r}{r_{g}}\right)\cos\theta\sin(\w t-\varphi+\phi_{0}) \, ,\\
        E_{A^\prime,r}&\approx0.3\e\a^{7/2}\Lambda\sin\theta\sin(\w t-\varphi+\phi_{0})\, ,\\
        E_{A^\prime,\theta}&\approx0.3\e\a^{7/2}\Lambda\cos\theta\sin(\w t-\varphi+\phi_{0})\, ,\\
        E_{A^\prime,\varphi}&\approx-0.3\e\a^{7/2}\Lambda\cos(\w t-\varphi+\phi_{0})\, .
    \end{aligned}
    \label{eq:EB_near}
\end{equation}
Here, $B_{A^\prime,r}$ is zero since the axial symmetry of $\vec{J}'$ renders itself curl-free in the radial direction. The characteristic $\a$ scalings of $E_{A'}$ and $B_{A'}$, {\it i.e.}, $B_{A^\prime}\sim \a^2 E_{A^\prime}\propto \a^{11/2}$, arise from the cancellation of lower-order terms in the small $\a$ expansion, different from those at $r\sim r_c$. $\Lambda=1/(r_{g}\sqrt{\pi G})$ is a factor characterising the DP-induced EM field strength for a given gravitational atom. Its value is solely determined by the BH mass and is equal to $7.3\times 10^9$G or $2.2\times 10^{14} \text{V}/\text{m}$ for the M87$^\ast$, where $M\approx 6.5 \times 10^9\,M_\odot$, and to $1.2\times 10^{13}\text{G}$ or $3.6\times 10^{17} \text{V/m}$ for the Sgr~A$^*$, where $M \approx 4.0 \times 10^6 M_\odot$. 

The BH accretion disk is not at rest in the observer frame. It is convenient to denote the EM field strength in the plasma frame as $\vec {\mc{E}}$ with $\mc{E}=|\vec{\mc{E}}|$ and $\vec {\mc{B}}$ with $\mc{B}=|\vec{\mc{B}}|$. Then in the direction transverse to the boost, the DP-induced components $\vec{\mc{E}}_{A^\prime\bot}$ and $\vec{\mc{B}}_{A^\prime\bot}$ are related to $\vec{B}_{A^\prime\bot}$ and $\vec{E}_{A^\prime\bot}$ through Lorentz transformation 
\begin{equation}
    \begin{aligned}
    \vec{ \mc{E}}_{A^\prime\bot}& =\c_{\rm p}(\pm \b_{\rm p}\vec{B}_{A^\prime\bot}+\vec{E}_{A^\prime\bot}) \,  ,\\
    \vec{\mc{B}}_{A^\prime\bot}& = \c_{\rm p}(\mp\b_{\rm p}\vec{E}_{A^\prime\bot}+\vec{B}_{A^\prime\bot}) \,  ,
    \end{aligned}
    \label{eq:LT}
\end{equation}
where $\b_{\rm p}$ and $\c_{\rm p}$ are  plasma velocity and boost factor in the observer frame. The EM components $\vec{\mc{E}}_{A^\prime\parallel}$ and $\vec{\mc{B}}_{A^\prime\parallel}$, which are parallel to the boost direction, keep identical in both reference frames. 


The DP-induced electric field at the BH disk could be strongly screened by plasma.
For the M87$^*$, its plasma temperature is $\sim 10^{11}\,$K and electron number density is $\sim 10^5\,\mathrm{cm}^{-3}$, leading to a Debye length of $\lambda_D \sim 10 - 100\,$m. The Sgr A$^*$ exhibits a similar plasma temperature, but its electron number density is higher~\cite{Chatterjee:2022pbo}, approximately $10^7~\mathrm{cm}^{-3}$, yielding a shorter Debye length of $\lambda_D \sim 1 - 10\,\mathrm{m}$. This implies that, in the M87$^*$ and Sgr A$^*$ plasma, the DP-induced electrostatic effect can only persist a  distance much shorter than the BH horizon and the geometric size of the surrounding plasma cloud, and thus be neglected. 
Instead, $\vec{\mc{B}}_{A^\prime}$ plays a key role in the DP polarimetry. 
The contributions to $\vec{\mc{B}}_{A^\prime}$ are primarily driven by $\vec{E}_{A^\prime}$, due to  $B_{A^\prime}\sim \a^2 E_{A^\prime}$ (see Eq.~(\ref{eq:EB_near})), despite the suppression of $\b_{\rm p} \sim\bigO(0.1)$ for both M87$^\ast$ and Sgr A$^*$.  
%
This is consistent with the assumptions taken for the ideal magnetohydrodynamics simulation in Ref.~\cite{Xin:2024trp}. Also, the potential synchrotron radiation and cascade production of electron-positron pairs caused by the  unscreened DP-induced electric field~\cite{Siemonsen:2022ivj} may not be efficient. 

The $\vec{\mc{B}}_{A^\prime}$ field can modify radiative transfer in the BH plasma in an oscillatory way, leaving an imprint in the time series of its polarimetric data. Generally, the polarimetric features of a radiation field are described by Stokes parameters $(I, Q, U, V)$, where the linear polarization PA is given by $\phi_{\rm PA}\equiv\frac{1}{2}\arg(Q+i\uu U)$. These parameters evolve during light propagation according to the radiative transfer equation~\cite{2011ApJ...737...21L,marszewski2021updated}:
\begin{align}
    \frac{d}{ds} 
    \begin{pmatrix}
        I \\
        Q \\
        U \\
        V
    \end{pmatrix}
    = 
    \begin{pmatrix}
        j_I \\
        j_Q \\
        j_U \\
        j_V
    \end{pmatrix}
    -
    \begin{pmatrix}
        \a_{I} & \a_{Q} & \a_{U} & \a_{V} \\
        \a_{Q} & \a_{I} & \rho_{V} & \rho_{U} \\
        \a_{U} & -\rho_{V} & \a_{I} & \rho_{Q} \\
        \a_{V} & -\rho_{U} & -\rho_{Q} & \a_{I}
    \end{pmatrix}
    \begin{pmatrix}
        I \\
        Q \\
        U \\
         V
    \end{pmatrix},
    \label{eq:rad_trans}
\end{align}
where $j_{I,Q,U,V}$ and $\a_{I,Q,U,V}$ are the plasma emissivity and absorption coefficients, respectively, and $\rho_{I,Q,U,V}$ are its Faraday mixing coefficients. For the BH environment, synchrotron radiation is highly relevant. We can align the $U$ parameter with the magnetic field then such that $j_U=\a_{U}=\rho_{U}=0$~\cite{Dexter:2016cdk}. The remaining coefficients are subject to $\vec{\mc{B}}_{A^\prime}$'s influence through their dependence on the $\vec{\mc{B}}$ field~\cite{marszewski2021updated}. Therefore, FR and FC, measured by $\rho_V$ and $\rho_Q$, along with light intensity, mediated by plasma emission and absorption, all could serve as observables with an oscillating signal signature.

Let us consider the weak-$\vec{\mc{B}}$ limit, where the BH radiation frequency $\nu \gg \nu_{\mc{B}} = \frac{e\mc{B}}{2\pi m_e}$, a condition applicable to most of the parameter space under exploration. For thermal plasma in this limit, the $j$, $\a$ and $\rho$ coefficients with a non-``$U$'' subscript are proportional to the plasma electron density $n_{e}$~\cite{Dexter:2016cdk}. The emission coefficients contain a factor of $\exp[-(\frac{27\nu}{4\nu_c})^{1/3}]$, 
where $\nu_{c}=3\sin\theta_{\mc{B}}\theta^2_e\nu_{\mc{B}}/2$ is a characteristic frequency. $\theta_{e}=T/m_{e}$ is a  normalized temperature parameter and $\theta_{\mc{B}}$ denotes the angle between the light wave vector and the magnetic field direction. The absorption coefficients are determined by Kirchoff's Law $\a_{I,Q,U,V} = f_\nu^{-1} j_{I,Q,U,V}$, where $f_{\nu}=\nu^3/(e^{\nu/T}-1)$ is blackbody function~\cite{Pandya:2016qfh}. Thus, these coefficients are exponentially suppressed for a weak $\vec{\mc{B}}$ field. Moreover, the FR coefficient $\rho_{V}$ is linear in $\mc{B}$, up to a mild logarithmic $\mc{B}$ term, and has a temperature dependence of $\propto\theta^{-2}_{e}\log\theta_{e}$ for $\theta_{e}\gg 1$. The FC  rotates between linear and circular polarizations of light. Its parameter $\rho_{Q}$ is $\propto \mc{B}^2\theta_{e}$, and for the relativistic plasmas, typical for the supermassive BHs, can be greatly enhanced~\cite{Dexter:2016cdk}. 



\section{Polarimetric Constraints on Photon - Dark Photon Kinetic Mixing}
Next, we will apply the polarimetry for detecting the ultralight DP, using the M87$^\ast$ and Sgr A$^*$ polarimetric data from the EHT. We track the evolution of the Stokes parameters within the DP cloud with the public relativistic polarised radiative transport code {\tt IPOLE}~\cite{Moscibrodzka:2017lcu,Noble:2007zx,ipoleUIUC}, by incorporating the DP-induced oscillating component into the EM field, and model the BH environmental electron number density with the radiatively inefficient accretion flow~\cite{pu2018probing}, namely 
\begin{align}
    n_{e}=n_{e}^0 \left(\frac{r+r_{+}}{r_g}\right)^{-1.1}e^{-  \frac{\cot^2 \theta}{2H^2}} \, .
    \label{eq:riaf}
\end{align}
Here, we set $n_{e}^0=3.4\times 10^5/\text{cm}^3$ and $3.0\times 10^7/\text{cm}^3$ for the M87$^*$ and the Sgr A$^*$, respectively, and $H=0.3$ for both, ensuring that the image intensity predicted by {\tt IPOLE} matches with the EHT observations~\cite{EHT-M87-8,EHT-SgrA-7}, with the effects of the DP cloud turned off. Note that in the {\tt IPOLE} the radiative transfer coefficients are computed using the public {\tt SYMPHONY} code~\cite{Dexter:2016cdk}, which employs numerical fitting functions assuming $\nu\gg\nu_{\mc{B}}$. Finally, the photon-DP mixing parameter can be constrained through its posterior probability distribution 
\begin{eqnarray}
&& P(\log_{10}\e\vert \{  \mathcal{O}_q   \}, \mu)  \nonumber \\ &=& \int_0^{2\pi}\frac{d\phi_0}{2\pi} \frac{1}{C} \prod_{q} \exp\left(-\frac{ - ( \mathcal{O}_q -  \tilde { \mathcal{O}}_q  )^2}{2 \sigma_q^2} \right) \, , \label{eq:post}
\end{eqnarray}
where $\mathcal{O}_q$ and $\tilde{\mathcal{O}}_q$ represent the data and predicted signal, respectively, $\sigma_q$ is the variance, and $q$ 
runs over all observations which are assumed to be independent. Moreover, uniform prior probability distributions have been implicitly assumed for $\log_{10}\e$, to ensure the normalization factor $ C $ to be a constant, and for $ \phi_0 $ which is marginalized by integrating over $[0, 2\pi]$. 

Let us consider the FR observation at the EHT first. Four polarimetric images of the M87$^\ast$ were captured on April 5, 6 and April 10, 11 of 2017, respectively~\cite{EHT-M87-7}. For sensitivity analysis, we introduce a universal sky plane for all M87$^\ast$ images using a polar coordinate system.  $(\widetilde{r},\widetilde{\varphi})$. 
The pole is positioned at the centre of the M87$^\ast$ image, with the polar axis aligned with the spin projection of M87$^\ast$ onto the plane. The intensity-weighted (IW) average PA for the $a$-th image, sorted chronologically, is defined as
\begin{align}
    \langle\phi_{\rm PA}^{\rm IW}(\widetilde{\varphi},t)
    \rangle_i^a =\frac{1}{2}\arg\left(\langle Q\times I (\widetilde\varphi,t)\rangle_i^a  + i \langle U\times I (\widetilde\varphi,t)\rangle_i^a \right),
    \label{eq:avg_chi}
\end{align}
where the average is performed over its $i$-th angular bin~\cite{EHT-M87-7}, with a width of $10^\circ$. We can define an observable then to measure its variation over two successive days (from 5 to 6 and 10 to 11 in April), yielding    
\begin{equation}
\begin{aligned}
    & \mathcal{O}_{q=\{ \tilde a, i \}} \equiv \D\langle\phi_{\rm PA}^{\rm IW} (\widetilde\varphi,t)\rangle_i^{\tilde a}  \\ & =\int_{t_{\rm obs}}\dfrac{dt}{t_{\rm obs}}\left(\langle\phi_{\rm PA}^{\rm IW}(\widetilde\varphi, t)\rangle_i^{2\tilde {a}-1}-\langle\phi_{\rm PA}^{\rm IW}(\widetilde\varphi, t)\rangle_i^{2\tilde {a}}\right) \, ,
    \label{eq:ob1}
\end{aligned}
\end{equation}
where $\tilde {a}=1,\,2$ denotes the $\tilde a$-th pair of images and $t_{\rm obs}=6$ hours accounts for the EHT's imaging time on each observation day (note that the de facto imaging time on each day differs slightly, and the uniform choice for $t_{\rm obs}$ is taken for convenience). This is designed to mitigate the impacts of temporal variation of the accretion disk over longer time scales.  
The analysis is restricted to the bins that are approximately Gaussian~\cite{chen2021stringent}, to ensure the applicability of Eq.~(\ref{eq:post}).

For the Sgr A$^*$, the EHT reported its full-image-averaged PA $\langle\phi_{\rm PA}(\widetilde\varphi,t)\rangle$, together with the uncertainties, for five nights~\cite{johnson2015resolved}. The data reveal a significant variability in the polarization PA across the observation period, suggesting an unstable plasma environment and dynamic magnetic field structure around the Sgr A$^*$. 
So we use the data from the observation day only exhibiting the most stable polarization feature for analysis, which were taken at the time $t_i$, with $i = 1$, 2, $\ldots$, 33. We then define an  observable 
\begin{equation}
\begin{aligned}
     \mathcal{O}_{q= i} \equiv  \D \langle\phi_{\rm PA}(\widetilde\varphi,t_i)\rangle  =\langle\phi_{\rm PA}(\widetilde\varphi, t_i)\rangle - \phi_{\rm PA}^n \, ,
    \label{eq:ob2}
\end{aligned}
\end{equation}
where $\phi_{\rm PA}^n$ denotes the actual noise level. Approximately, we assume $\phi_{\rm PA}^n$  to be constant for the considered observation day, with a uniform prior distribution over $[0, \pi]$, and marginalize it away in the analysis. The observables in Eq.~(\ref{eq:ob1}) and Eq.~(\ref{eq:ob2}) both have an variance $\sigma_q \sim \mathcal O(10)^\circ$.

\begin{figure}[!htbp]
    \centering
    \includegraphics[width=1.05\columnwidth]{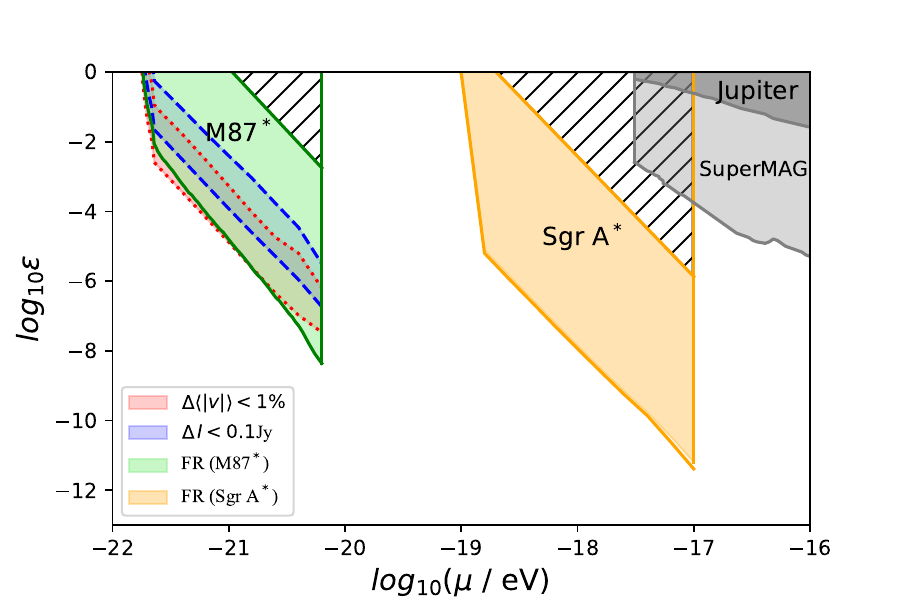}
    \caption{95\% confidence level (CL) exclusion limits on the photon-DP kinetic mixing parameter $\e$, using the EHT polarimetric images of the M87$^*$ and Sgr A$^*$. For the M87$^*$, the green shaded region is excluded by the FR imaging. The blue and red shaded regions are excluded by the intensity and dichroism imaging. 
    In the hatched region, the {\tt IPOLE} calculation is unreliable because the assumption of $\nu\gg\nu_{\mc{B}}$ 
    gets broken. 
    For the Sgr A$^*$, the orange shaded region is excluded by the FR imaging. As a reference, we also include the exclusion limits obtained from the Jupiter~\cite{Yan:2023kdg} and SuperMAG~\cite{Fedderke:2021rrm}. }   
    \label{fig:result}
\end{figure}

Given these EHT images, we demonstrate in Fig.~\ref{fig:result} the 95\% CL FR limits on the photon-DP kinetic mixing parameter $\e$. 
The analysis is restricted to $\a<0.3$ to meet the superradiance condition for the considered DP cloud mode of $(n,j,l,m)=(1,1,0,1)$~\cite{Dolan:2007mj}. These limits approximately scale with the DP mass as $\sim \mu^{7/2}$ in the relevant mass range. For $\nu \gg \nu_{\mc{B}}$, the DP-induced FR is $\sim \int dl\uu n_e \mc{B}_{A^\prime}$, with the integration performed along the light trajectory. Since the integrand is proportional to $\mc{B}_{A^\prime} \propto E_{A^\prime} \propto \varepsilon \a^{7/2} $, such a scaling behavior is naturally expected. 
Near the low DP mass end, the exclusion curves are both subject to a turning point, to the left of which the DP cloud fails to achieve full density during cosmic age, further weakening the limits. Notably, the electron density near the Sgr A$^*$ is higher than that near the  M87$^*$ by orders of magnitude~\cite{Chatterjee:2022pbo}, resulting in the limits from the Sgr A* much stronger than those from the M87$^*$.   


Besides $Q$ and $U$, the Stokes parameters $I$ and $V$ can also be utilized to probe for the DP, where the DP-induced magnetic field contributes through f radiation and FC, respectively. As a demonstration, let us consider two EHT observables for the M87$^\ast$ only. One is total compact flux density of the images $F_{\rm cpct}$ and it has been measured with a precision of $\sim 0.1\,$Jy at the EHT~\cite{EHT-M87-4}. Another one is average fractional circular polarization $\langle\lvert V \rvert\rangle=\int d\sigma\, \lvert V\rvert/\int d\sigma \,\lvert I\rvert$, where $\int d\sigma$ denotes an integration over the image. The EHT precision of measuring $\langle\lvert V \rvert\rangle$ is currently $\sim 10^{-2}$~\cite{EHT-M87-9}. However, we found no temporal information over the observation period reported in these literatures. To estimate the sensitivities of these two observables, we thus require the $\mc{B}_{A^\prime}$-induced oscillation amplitude in these two observables to be smaller than their measurement precisions. The excluded parameter regions are also displayed in Fig.~\ref{fig:result}, as blue and red shaded bands, respectively. We emphasize that, due to the absence of temporal variance as a signal signature in the analysis, these results should be interpreted as ``projected limits'' rather than ``actual limits'', based on the current EHT precision level.

These exclusion limits of $\varepsilon$ demonstrate a feature of scaling with $\mu^{7/2}$, consistent with our expectation. However, the exclusion power of $F_{\rm cpct}$ and $\langle\lvert V \rvert\rangle$ become weak for a strong $\vec{\mc{B}}_{A'}$ field, yielding a relatively narrow exclusion band in both cases. This outcome is caused by the variation of $|dI/d\mc{B}|$ and $|dV/d\mc{B}|$ as the $\mc{B}_{A'}$ or the $\mc{B}$ increases.  
As a qualitative discussion, we can turn off the Faraday mixing coefficients and decouple the $V$ parameter from the radiative transfer equation. Then for homogeneous and thermal plasma~\cite{1991epdm.book.....M} we have $I \approx f_\nu\left[2-\exp(-(\a_I+\a_Q)L)-\exp(-(\a_I-\a_Q)L)\right]$, where $L$ is plasma thickness. Then we can find that $|dI/d\mc{B}|$ is  $\propto \mc{B}^{-4/3}$ for weak $\mc{B}$, and exponentially suppressed for $(\a_I\pm\a_Q)L\gtrsim 1$ where $\mc{B}$ becomes relatively strong and $I$ gets saturated. Similarly, the FC  parameter $\rho_Q$, which is key for rotating the $U$ parameter to the $V$ parameter, is proportional to $\mc{B}^2$ for small $\mc{B}$ while becomes exponentially suppressed as $\mc{B}$ becomes large. The exclusion band obtained from measuring $\langle\lvert V \rvert\rangle$ thus gets explained. So, these polarimetric observables may have comparable sensitivities at the EHT in probing for the photon-DP mixing parameter when this parameter is relatively small or the DP-induced $\vec{\mc{B}}_{A'}$ is relatively weak.    

\section{Summary and Outlook}
In this Letter, we propose and develop DP polarimetry, leveraging the effect of FR particularly, to detect the DP. We demonstrate this method by searching for ultralight DPs produced through superradiance of supermassive BHs. Then using the polarimetric images of the M87$^*$ and Sgr A$^*$ from the EHT, we show that novel limits on the photon-DP kinetic mixing parameter can be set for the DP with a mass within $10^{-22} - 10^{-20}\,$eV and $10^{-19} - 10^{-17}\,$eV, respectively. These mass ranges have been rarely explored before using a non-gravitational method, thereby complementing this work with existing studies greatly. Moreover, we show that the Stokes parameters $I$ and $V$ could be also applied to probe for the DP through the effects of synchrotron radiation and FC.

Following this proof-of-concept investigation, we can readily recognize several important directions for next-step explorations. 
For example, we can apply the DP polarimetry for the DP detection in the laboratory polarimetric experiments or other astronomical and cosmological polarimetric observations. We can also expand the study to axion, which can induce an effective current through the axion Chern-Simons coupling in the Maxwell's equations. Different from its DP counterpart, this current is proportional to the external magnetic field strength. We leave these explorations to future work.

{\bf Note Added:} When this Letter was in finalization, the paper~\cite{Acevedo:2025ysl} appeared on arXiv. Although this paper aims to address the detection of the ultralight DP also, it employs a completely different method based on the DP interactions with charged particles (inverse Compton scattering).  

\begin{acknowledgments}
\subsection*{Acknowledgments}
We would thank Yifan Chen, Shuo Xin and Yue Zhao for useful discussions, and the Center for High Throughput Computing at the University of Wisconsin-Madison for providing computing resources~\cite{https://doi.org/10.21231/gnt1-hw21}. J.R. is supported in part by the National Natural Science Foundation of China under Grant No. 12275276. 
\end{acknowledgments}

\bibliography{SMBH}

\end{document}